\PassOptionsToClass{a4paper}{article}
\documentclass{svproc}
\usepackage[papersize={155mm,235mm},textwidth=122mm,textheight=193mm,headsep=12pt,centering]{geometry}
\usepackage{hyperref}
\usepackage{url}

\usepackage{graphicx}
\usepackage{bm}
\usepackage[font=footnotesize]{subfig}
\usepackage{amsmath}
\usepackage{xcolor}
\usepackage{eso-pic}

\graphicspath{{./figures/}}

\newcommand{\Vector}[1]{\bm{#1}}
\newcommand{\Matrix}[1]{\bm{#1}}
\newcommand{\Transpose}{\mathrm{T}}
\newcommand{\refEq}[1]{(\ref{#1})}               
\newcommand{\refFig}[1]{Fig.~\ref{#1}}           
\newcommand{\refSec}[1]{Section~\ref{#1}}        
\newcommand{\Tsample}{T_\mathrm{s}}

\begin{document}

\AddToShipoutPicture*{%
  \AtPageUpperLeft{%
    \setlength\unitlength{1cm}%
    \put(0,-0.9){\begin{minipage}[c]{\paperwidth}
    \footnotesize\centering\textcolor{black!50}{%
    This is a preprint of an article presented at the \emph{22nd Polish Control Conference (PCC2026)}.\\
    The final authenticated version is available online at:} \ \textcolor{blue!60}{\url{https://doi.org/10.1007/978-3-032-32210-4_22}}%
    \end{minipage}}%
  }
}

\mainmatter

\title{Downgrading ADRC to a\\ Glorious Proportional Controller}
\titlerunning{Downgrading ADRC to a Glorious Proportional Controller}

\author{Gernot Herbst
}
\authorrunning{Gernot Herbst}
\tocauthor{Gernot Herbst}

\institute{%
    University of Applied Sciences Zwickau, 08056 Zwickau, Germany\\
    \email{gernot.herbst@whz.de}%
}

\maketitle

\begin{abstract}
    This article parameterizes discrete-time error-based ADRC to obtain P controller behavior at runtime, reveals the limits of this approach, and simplifies its structure while retaining the ADRC features of delay compensation and output limitation.
    For integrator plants, the resulting discrete-time P controller can be tuned using existing ADRC tuning equations up to dead-beat dynamics even under these conditions, and will be shown to be equivalent to a Smith predictor approach.
    \keywords{ADRC, P controller, saturation, delay compensation}
\end{abstract}
%

\section{Introduction and Main Idea}

Many PID controllers allow to parameterize for pure P or I behavior at runtime.
For active disturbance rejection control (ADRC), the dynamics are seemingly fixed by the controller/observer structure.
For an inner loop of a cascade control system, and especially for integrator plants such as the current control loop of power electronics converters, a simple proportional controller may be sufficient, even if a more sophisticated one has already been deployed.

This article will address two questions:
To what extent can ADRC replicate P behavior only through parameter choice?
And is it possible to derive a ``better'' P controller inheriting advanced features of ADRC, especially the handling of dead times and controller saturation by trimming down its structure to the essentials?
As there are practically no references for this very special undertaking, let us directly get started, pointing to foundational literature as needed on the way.

Reducing ADRC to a P controller is not possible with every variant.
In this article, the technical basis will be linear error-based ADRC.
After summarizing its most important equations in \refSec{sec:eADRC} (focusing only on the practically relevant discrete-time form), this choice will be clear:
One can select poles and zeros of its transfer function equivalent to achieve P controller behavior.

In \refSec{sec:Parameterization} its parameters will be used to obtain P behavior in a step-by-step approach, revealing what is possible, where modifications are needed, and discussing edge cases.
Learning from these aspects, the original ADRC structure will then be reduced to the required essentials in \refSec{sec:FinalP}.
Especially with an integrator plant, the resulting controller will allow easy tuning up to achieving dead-beat dynamics even under the real-world restrictions of limited control signals and having delays in the loop.
As a final surprise, it will turn out that the resulting controller is equivalent to a Smith predictor in that case.

\section{Summary of Discrete-Time Error-Based ADRC}
\label{sec:eADRC}

\begin{figure}
    \subfloat[State space form]{%
        \includegraphics[scale=0.6]{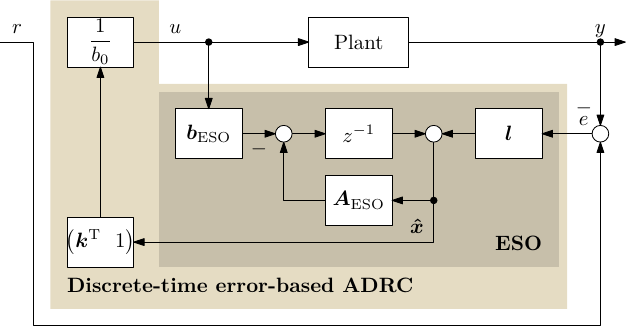}
        \label{fig:DT_eADRC_SS}
    }%
    \hfill%
    \subfloat[Transfer function equivalents]{%
        \begin{minipage}[b]{0.4\textwidth}
            \includegraphics[scale=0.6]{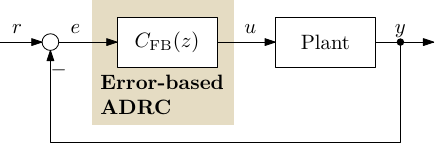}\\[1em]
            \includegraphics[scale=0.5]{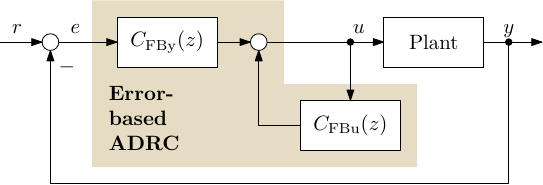}
        \end{minipage}
        \label{fig:DT_eADRC_TF}
    }%
    \caption{%
        Control loop with error-based ADRC given in different equivalent forms
    }
    \label{fig:DT_eADRC}
\end{figure}

Since most real-world implementations of ADRC would be software-based, typically on programmable logic controllers (PLCs) or microcontrollers (MCUs), its discrete-time error-based variant will now be summarized.
Very similar to the ``traditional'' (output-based) variant introduced in \cite{Gao:2003}, it consists of an extended state observer (ESO) and an outer state-feedback controller, cf.\ \refFig{fig:DT_eADRC_SS}.
The main differences are that (1) the control law depends only on the observer states (and not on the reference signal $r$), and (2) the external input to the observer is the control error $e$ (and not the controlled variable $y$):
\begin{gather}
    u(k) = \frac{1}{b_0} \cdot \begin{pmatrix} \Vector{k}^\Transpose & 1 \end{pmatrix}
    \cdot \Vector{\hat{x}}(k)
    ,
    \label{eqn:ADRC_DTe_SS_Controller}
    \\
    \Vector{\hat{x}}(k) = \Matrix{A}_\mathrm{ESO} \cdot \Vector{\hat{x}}(k-1) - \Vector{b}_\mathrm{ESO} \cdot u(k-1) + \Vector{l} \cdot e(k)
    .
    \label{eqn:ADRC_DTe_SS_Observer}
\end{gather}

As shown in \cite{Madonski:2023} for the continuous-time case, its behavior can be represented using a single transfer function, the feedback controller $u(z) = C_\mathrm{FB}(z) \cdot e(z)$ that it shares with output-based ADRC:
\begin{equation}
    C_\mathrm{FB}(z)
    = \frac{1}{b_0} \cdot \begin{pmatrix} \Vector{k}^\Transpose & 1 \end{pmatrix}
    \cdot \left(
    \Matrix{I} - z^{-1} \cdot \left( \Matrix{A}_\mathrm{ESO} - \frac{1}{b_0} \cdot          \Vector{b}_\mathrm{ESO} \cdot \begin{pmatrix} \Vector{k}^\Transpose & 1 \end{pmatrix} \right)
    \right)^{-1}
    \cdot \Vector{l}
    .
    \label{eqn:ADRC_DT_CFB}
\end{equation}

In \refEq{eqn:ADRC_DTe_SS_Observer}, zero-order hold discretization with a sampling interval $\Tsample$ and the current observer approach have been used, cf.\ \cite{Miklosovic:2006}.
The results were given in the compact notation as in \cite{Herbst:2025}, where the details are:
\begin{gather}
    \Matrix{A}_\mathrm{ESO} = \Matrix{A}_\mathrm{d} - \Vector{l} \cdot \Vector{c}^\Transpose_\mathrm{d} \cdot \Matrix{A}_\mathrm{d}
    \quad\text{and}\quad
    \Vector{b}_\mathrm{ESO} = \Vector{b}_\mathrm{d} - \Vector{l} \cdot \Vector{c}^\Transpose_\mathrm{d} \cdot \Vector{b}_\mathrm{d}
    \label{eqn:ADRC_DT_AESO_BESO}
    \\
    \text{with}\quad
    \Matrix{A}_\mathrm{d} = \Matrix{I} + \sum_{i = 1}^\infty \displaystyle\frac{\Matrix{A}^i \Tsample^i}{i!}
    ,\quad
    \Vector{b}_\mathrm{d} = \left( \sum_{i = 1}^\infty \displaystyle\frac{\Matrix{A}^{i-1} \Tsample^i}{i!} \right) \cdot \Vector{b}
    ,\quad
    \Vector{c}^\Transpose_\mathrm{d} = \Vector{c}^\Transpose
    .
    \label{eqn:ADRC_DT_AdBdCd}
\end{gather}

For first-order ADRC, which is built around a model of a disturbed first-order integrator $b_0 / s$, the continuous-time as well as resulting discrete-time observer matrices and vectors in \refEq{eqn:ADRC_DT_AdBdCd} and \refEq{eqn:ADRC_DT_AESO_BESO} are:
\begin{gather}
    \Matrix{A} = \begin{pmatrix}
        0 \  &  \ 1 \\[0.5ex]
        0 \  &  \ 0
    \end{pmatrix}
    ,\quad
    \Vector{b} = \begin{pmatrix}
        b_0  \\[0.5ex]
        0
    \end{pmatrix}
    ,\quad\text{and}\quad
    \Vector{c}^\Transpose = \begin{pmatrix}
        1 \  &  \ 0
    \end{pmatrix}
    ,
    \label{eqn:ADRC_CT_ABC}
    \\
    \Matrix{A}_\mathrm{ESO} = \begin{pmatrix}
        1 - l_1 \  &  \ \Tsample - l_1 \Tsample  \\[0.5ex]
        - l_2 \  &  \ 1 - l_2 \Tsample
    \end{pmatrix}
    \quad\text{and}\quad
    \Vector{b}_\mathrm{ESO} = \begin{pmatrix}
        b_0 \Tsample - l_1 b_0 \Tsample  \\[0.5ex]
        - l_2 b_0 \Tsample
    \end{pmatrix}
    .
    \label{eqn:ADRC1_DT_AESO_BESO}
\end{gather}

Dedicated discrete-time tuning is necessary at least for higher dynamic requirements \cite{Herbst:2025}.
Desired poles for the closed control loop as well as the observer are obtained by mapping from $s$-plane to $z$-plane locations.
In case of bandwidth tuning \cite{Gao:2003} with bandwidths $\omega_\mathrm{CL}$ and $\omega_\mathrm{ESO}$ the resulting equations are:
\begin{gather}
    k_1 = \frac{1 - z_\mathrm{CL} }{ \Tsample }
    ,\quad
    l_1 = 1 - z_\mathrm{ESO}^2
    ,\quad\text{and}\quad
    l_2 = \frac{1}{\Tsample} \cdot \left(1 - z_\mathrm{ESO}\right)^2,
    \label{eqn:ADRC1_DT_KL_BWTuning}
    \\
    \text{with}\quad
    z_\mathrm{CL} = \mathrm{e}^{-\omega_\mathrm{CL} \Tsample}
    \quad\text{and}\quad
    z_\mathrm{ESO} = \mathrm{e}^{-\omega_\mathrm{ESO} \Tsample}
    .
    \label{eqn:ADRC_DT_Poles}
\end{gather}

As described in detail in \cite{Herbst:2025}, inserting \refEq{eqn:ADRC1_DT_AESO_BESO} and \refEq {eqn:ADRC1_DT_KL_BWTuning} in \refEq{eqn:ADRC_DT_CFB} leads to $C_\mathrm{FB}(z)$:
\begin{equation}
    C_\mathrm{FB}(z)
    = \frac{ u(z) }{ e(z) }
    = \frac{ \beta_0 + \beta_1 z^{-1} }{ 1 + \alpha_1^\mathrm{FB} z^{-1} } \cdot \frac{1}{1 - z^{-1}}
    ,
    \label{eqn:DT_eADRC_TF}
\end{equation}
with general and bandwidth-tuning based expressions for the coefficients being:
\begin{align}
    \alpha_1^\mathrm{FB}
    &= (\Tsample k_1 - 1) \cdot (1 - l_1)
    = -z_\mathrm{CL} z_\mathrm{ESO}^2
    ,
    \label{eqn:ADRC1_DT_CFB_alpha1}
    \\
    \beta_0
    &= \frac{ k_1 l_1 + l_2 }{ b_0 }
    = \frac{ z_\mathrm{CL} z_\mathrm{ESO}^2 - 2 z_\mathrm{ESO} - z_\mathrm{CL} + 2 }{ b_0 \Tsample }
    ,\quad\text{and}
    \label{eqn:ADRC1_DT_CFB_beta0}
    \\
    \beta_1
    &= \frac{ \Tsample k_1 l_2 - k_1 l_1 - l_2 }{ b_0 }
    = \frac{ 2 z_\mathrm{CL} z_\mathrm{ESO} - 2 z_\mathrm{CL} z_\mathrm{ESO}^2 + z_\mathrm{ESO}^2 - 1 }{ b_0 \Tsample }
    .
    \label{eqn:ADRC1_DT_CFB_beta1}
\end{align}

The transfer function \refEq{eqn:DT_eADRC_TF} can either be seen as a PI controller with a low-pass filter, or an integral controller with an additional lead-lag element in series.
The structure of \refEq{eqn:DT_eADRC_TF} already suggests a possible solution to achieve P controller behavior: remove a pole by setting $\alpha_1^\mathrm{FB} = 0$ and select the $\beta$ coefficients to cancel the integrator pole---more on that in \refSec{sec:Parameterization}.

Without derivation, which can also be found in \cite{Herbst:2025}, a second (so-called ``dual feedback'') transfer function representation is possible.
As visible in \refFig{fig:DT_eADRC_TF}, it maintains the explicit feedback path of the control signal back into the observer as in the state-space form.
This is important for possible ADRC extensions taking control signal saturation and delay into account:
\begin{equation}
    u(z) = C_\mathrm{FBy}(z) \cdot e(z) + C_\mathrm{FBu}(z) \cdot u(z)
    \quad\rightarrow\quad
    u(z) = \underbrace{ \frac{ C_\mathrm{FBy}(z) }{ 1 - C_\mathrm{FBu}(z) } }_{ C_\mathrm{FB}(z) } \cdot e(z)
    .
    \label{eqn:ADRC_DT_DFTF}
\end{equation}

Both representations will be used here: $C_\mathrm{FB}(z)$ for assessment of the dynamic behavior and easy comparison against the target P controller, and the ``dual feedback'' form to integrate add-on functionality.
The first-order case yields:

\begin{equation}
    C_{\mathrm{FBy}}(z) = \frac{\beta_0 + \beta_1 z^{-1}}{1 + \alpha_1 z^{-1} + \alpha_2 z^{-2}}
    \quad \text{and} \quad
    C_{\mathrm{FBu}}(z) = \frac{ z^{-1} \cdot \left( \gamma_0 + \gamma_1 z^{-1} \right) }{1 + \alpha_1 z^{-1} + \alpha_2 z^{-2}}
    \label{eqn:ADRC1_DT_DFTF}
\end{equation}
with the $\beta_0$ and $\beta_1$ parameters being the same as in \refEq{eqn:ADRC1_DT_CFB_beta0} and \refEq{eqn:ADRC1_DT_CFB_beta1}, as well as
\begin{align}
    \alpha_1
    &= l_1 + \Tsample l_2 - 2 = -2 z_\mathrm{ESO}
    ,
    \label{eqn:ADRC1_DFTF_DFTF_alpha1}
    \\
    \alpha_2
    &= 1 - l_1 = z_\mathrm{ESO}^2
    ,
    \label{eqn:ADRC1_DFTF_DFTF_alpha2}
    \\
    \gamma_0
    &= \Tsample \cdot (- k_1 + k_1 l_1 + l_2)
    = z_\mathrm{CL} z_\mathrm{ESO}^2 - 2 z_\mathrm{ESO} + 1
    ,\quad\text{and}
    \label{eqn:ADRC1_DT_DFTF_gamma0}
    \\
    \gamma_1
    &= \Tsample k_1 \cdot (1 - l_1)
    = z_\mathrm{ESO}^2 - z_\mathrm{CL} z_\mathrm{ESO}^2
    .
    \label{eqn:ADRC1_DT_DFTF_gamma1}
\end{align}

\section{Parameterization of Error-Based ADRC for P Control}
\label{sec:Parameterization}

The first option to be examined will only use the parameter interface of first-order discrete-time error-based ADRC.
This route will allow to achieve proportional behavior in a loop where this controller type has been implemented---similar to using only the P component of a PID controller.

    \subsection{Filtered P Controller}
    \label{sec:FilteredP}

        \subsubsection{Canceling the Integrator Pole}
%
%
in the controller transfer function $C_\mathrm{FB}(z)$ is the essential step to achieve proportional behavior, albeit still with a low-pass filter component active.
From \refEq{eqn:DT_eADRC_TF} it becomes obvious that $\beta_1 = -\beta_0$ must hold:
\begin{equation}
    C_\mathrm{FB}(z)
    = \frac{ \beta_0 + \beta_1 z^{-1} }{ 1 + \alpha_1^\mathrm{FB} z^{-1} } \cdot \frac{1}{1 - z^{-1}}
    \stackrel{{\color{orange}\beta_1 = -\beta_0}}{=}
    \frac{ \beta_0 }{ 1 + \alpha_1^\mathrm{FB} z^{-1} } \cdot \frac{ 1 - z^{-1} }{ 1 - z^{-1} }
    = \frac{ \beta_0 }{ 1 + \alpha_1^\mathrm{FB} z^{-1} }
    .
    \label{eqn:FilteredP_CFB_Approach}
\end{equation}

Inspection of \refEq{eqn:ADRC1_DT_CFB_beta0} and \refEq{eqn:ADRC1_DT_CFB_beta1} reveals that setting the controller gain to $k_1 = 0$ will achieve this goal ($\beta_1 = -\beta_0$).
If in the controller implementation $k_1$ is tuned by selecting a closed-loop pole $z_\mathrm{CL}$ or a bandwidth $\omega_\mathrm{CL}$, $k_1 = 0$ results from a pole placement at $z_\mathrm{CL} = 1$, i.\,e.\ $\omega_\mathrm{CL} = 0$.
The expressions for transfer function coefficients as well as controller and observer gains will then be:
\begin{align}
    &\beta_1
    = \frac{ 2 z_\mathrm{CL} z_\mathrm{ESO} - 2 z_\mathrm{CL} z_\mathrm{ESO}^2 + z_\mathrm{ESO}^2 - 1 }{ b_0 \Tsample }
    \stackrel{{\color{orange}z_\mathrm{CL} = 1}}{=}
    -\frac{ z_\mathrm{ESO}^2 - 2 z_\mathrm{ESO} + 1 }{ b_0 \Tsample }
    ,
    \\
    &\beta_0
    = \frac{ z_\mathrm{CL} z_\mathrm{ESO}^2 - 2 z_\mathrm{ESO} - z_\mathrm{CL} + 2 }{ b_0 \Tsample }
    \stackrel{{\color{orange}z_\mathrm{CL} = 1}}{=}
    \frac{ z_\mathrm{ESO}^2 - 2 z_\mathrm{ESO} + 1 }{ b_0 \Tsample }
    = -\beta_1
    ,
    \\
    &\alpha_1^\mathrm{FB}
    = -z_\mathrm{CL} z_\mathrm{ESO}^2
    \stackrel{{\color{orange}z_\mathrm{CL} = 1}}{=}
    -z_\mathrm{ESO}^2
    ,\quad\text{as well as}
    \label{eqn:FilteredP_CFB_alpha1}
    \\
    &k_1 = \frac{1 - z_\mathrm{CL} }{ \Tsample }
    \stackrel{{\color{orange}z_\mathrm{CL} = 1}}{=}
    0
    ,\quad
    l_1 = 1 - z_\mathrm{ESO}^2
    ,\quad\text{and}\quad
    l_2 = \frac{1}{\Tsample} \cdot \left(1 - z_\mathrm{ESO}\right)^2
    \label{eqn:FilteredP_CFB_K_L}
    .
\end{align}

        \subsubsection{Open-Loop Properties of the Filtered P Controller}
%
%
are found by comparing the controller transfer function \refEq{eqn:FilteredP_CFB_TF} (given both using observer gains and bandwidth parameterization) with generic first-order low-pass dynamics \refEq{eqn:DT_LowPass}:
\begin{equation}
    C_\mathrm{FB}(z)
    = \frac{ u(z) }{ e(z) }
    = \frac{ l_2 }{ b_0 \cdot \left( 1 + (l_1 - 1) \cdot z^{-1} \right) }
    = \frac{
        \left(1 - z_\mathrm{ESO}\right)^2
    }{
        b_0 \Tsample \cdot \left( 1 - z_\mathrm{ESO}^2 \cdot z^{-1} \right)
    }
    ,
    \label{eqn:FilteredP_CFB_TF}
\end{equation}
\begin{equation}
    G_\mathrm{LP}(s)
    = \frac{ K }{ 1 + \frac{1}{\omega_\mathrm{LP}} \cdot s }
    \quad \stackrel{\text{ZOH}}{\longrightarrow} \quad
    G_\mathrm{LP}(z)
    = \frac{
        K \cdot \left( 1 - \mathrm{e}^{-\omega_\mathrm{LP} \Tsample} \right) \cdot z^{-1}
    }{
        1 - \mathrm{e}^{-\omega_\mathrm{LP} \Tsample} \cdot z^{-1}
    }
    .
    \label{eqn:DT_LowPass}
\end{equation}

Ignoring the additional unit delay in the numerator one can draw a comparison of the approximate continuous-time behavior from the denominators, while the DC gain of the filtered P controller is obtained from the final value theorem:
\begin{gather}
    z_\mathrm{ESO}^2 = \mathrm{e}^{-2 \omega_\mathrm{ESO} \Tsample}
    \stackrel{!}{=}  \mathrm{e}^{-\omega_\mathrm{LP} \Tsample}
    \quad\rightarrow\quad
    \omega_\mathrm{LP} = 2 \omega_\mathrm{ESO}
    ,
    \\
    K = C_\mathrm{FB}(z \to 1)
    = \frac{
        \left(1 - z_\mathrm{ESO}\right)^2
    }{
        b_0 \Tsample \cdot \left( 1 - z_\mathrm{ESO}^2 \right)
    }
    .
    \label{eqn:FilteredP_DCGain}
\end{gather}

One possible tuning method for the filtered P controller could therefore consist of selecting a controller gain $K$ and filter bandwidth $\omega_\mathrm{LP}$, computing $z_\mathrm{ESO}$ with $\omega_\mathrm{ESO} = \frac{1}{2} \omega_\mathrm{LP}$, and then solving \refEq{eqn:FilteredP_DCGain} for $b_0$.

        \subsubsection{Closed-Loop Dynamics with the Filtered P Controller}
%
%
will now be examined with a integrator plant $P(s) = \frac{b}{s}$, as this is the canonical plant first-order ADRC is being built around.
With a zero-order hold behavior of the actuator, the discretized plant model is:
\begin{equation}
    P(z)
    = \frac{y(z)}{u(z)}
    = \frac{ b \cdot \Tsample \cdot z^{-1} }{ 1 - z^{-1} }
    .
    \label{eqn:DT_IntegratorPlant}
\end{equation}

Assuming perfect knowledge $b_0 = b$ for now, the closed-loop transfer function, given both in terms of observer gains and using bandwidth parameterization, is:
\begin{align}
    &G_\mathrm{CL}(z)
    = \frac{y(z)}{r(z)}
    = \frac{ C_\mathrm{FB}(z) \cdot P(z) }{ 1 + C_\mathrm{FB}(z) \cdot P(z) }
    \label{eqn:FilteredP_GCL_Integrator}
    \\
    &= \frac{
        l_2 \Tsample \cdot z^{-1}
    }{
        1 + (l_1 + l_2 \Tsample - 2) \cdot z^{-1} + (1 - l_1) \cdot z^{-2}
    }
    = \frac{
        (1 - z_\mathrm{ESO})^2 \cdot z^{-1}
    }{
        1 - 2 z_\mathrm{ESO} \cdot z^{-1} + z_\mathrm{ESO}^2 \cdot z^{-2}
    }
    .
    \notag
\end{align}

As to be expected, the closed loop has unity gain.
A more interesting result is that the closed-loop poles are exactly those designed for the observer!
Therefore existing design equations can be used, such as one given in \cite{Herbst:2025} relating bandwidth to the desired settling time:
\begin{equation}
    \omega_\mathrm{ESO} \approx \frac{ 6 }{ T_\mathrm{settle,98\%} }
    .
    \label{eqn:FilteredP_Design_Tsettle}
\end{equation}

        \subsubsection{Delay Compensation}
%
%
is a powerful ADRC feature which can be easily added to a solution.
If there is a delay of $D$ sampling intervals in the loop, a simple remedy is to feed $u(k-1-D)$ instead of $u(k-1)$ back to the observer \refEq{eqn:ADRC_DTe_SS_Observer}.
The resulting controller transfer function can be calculated from the ``dual feedback'' form \refEq{eqn:ADRC_DT_DFTF} by placing an additional $z^{-D}$ delay in the feedback path.
Since $k_1 = 0$ still holds, we have $\beta_0 = -\beta_1 = \frac{l_2}{b_0}$, $\gamma_0 = \Tsample l_2$, and $\gamma_1 = 0$ in \refEq{eqn:ADRC1_DT_DFTF} and obtain:
\begin{equation}
    C_\mathrm{FB}(z) = \frac{ C_\mathrm{FBy}(z) }{ 1 - C_\mathrm{FBu}(z) \cdot z^{-D} }
    = \frac{
        \beta_0 \cdot \left( 1 - z^{-1} \right)
    }{
        \left( 1 + \alpha_1 \cdot z^{-1} + \alpha_2 \cdot z^{-2} \right) - \gamma_0 \cdot z^{-1-D}
    }
    .
\end{equation}

To illustrate the effect of delay compensation, we enhance the integrator plant model \refEq{eqn:DT_IntegratorPlant} with an input delay $z^{-D}$:
\begin{equation}
    P(z)
    = \frac{y(z)}{u(z)}
    = \frac{ b \cdot \Tsample \cdot z^{-1-D} }{ 1 - z^{-1} }
    ,
    \label{eqn:DT_IntegratorPlantDelay}
\end{equation}
and compute the closed-loop transfer function, once more assuming $b_0 = b$:
\begin{align}
    &G_\mathrm{CL}(z)
    = \frac{
        \beta_0 b_0 \Tsample \cdot z^{-1-D}
    }{
        \beta_0 b_0 \Tsample \cdot z^{-1-D} + \left( 1 + \alpha_1 \cdot z^{-1} + \alpha_2 \cdot z^{-2} \right) - \gamma_0 \cdot z^{-1-D}
    }
    \notag\\
    &= \frac{
        \beta_0 b_0 \Tsample \cdot z^{-1-D}
    }{
        1 + \alpha_1 \cdot z^{-1} + \alpha_2 \cdot z^{-2}
    }
    = \frac{
        l_2 \Tsample \cdot z^{-1-D}
    }{
        1 + (l_1 + l_2 \Tsample - 2) \cdot z^{-1} + (1 - l_1) \cdot z^{-2}
    }
    \label{eqn:FilteredP_GCL_IntegratorDelay}
    ,
\end{align}
yielding exactly the same behavior as in \refEq{eqn:FilteredP_GCL_Integrator}, except for the unavoidable additional delay $z^{-D}$.
This is a very nice feature (not only) for a P controller, as one would otherwise have to explicitely decrease the controller gain when being faced with a reduced phase margin due to delays in the loop.

        \subsubsection{Controller Output Limitation}
%
%
is another topic where ADRC usually shines, as feeding back the limited controller output $u_\mathrm{lim}$ instead of $u$ into the observer \refEq{eqn:ADRC_DTe_SS_Observer} avoids windup effects.
While we would not have to expect windup in a proportional controller, this approach turns out to be a kind of Achilles' heel for the (filtered) P controller developed so far.
The reason is that the pole/zero pair at $z = 1$ caused by $\beta_0 = -\beta_1$ in $C_\mathrm{FB}(z)$ only cancels out in \refEq{eqn:FilteredP_CFB_Approach} when there is no saturation present, i.e., for the linear case:
\begin{equation}
    C_\mathrm{FB}(z)
    = \frac{ \beta_0 }{ 1 + \alpha_1^\mathrm{FB} z^{-1} } \cdot \frac{ 1 - z^{-1} }{ 1 - z^{-1} }
    \stackrel{{\color{orange}u_\mathrm{lim} = u}}{=}
    \frac{ \beta_0 }{ 1 + \alpha_1^\mathrm{FB} z^{-1} }
    .
\end{equation}

With a limited controller output one cannot simplify the ``dual feedback'' structure
\refEq{eqn:ADRC_DT_DFTF} to the $C_\mathrm{FB}(z)$ we aimed at, but is stuck with:
\begin{equation}
    u(z) = C_\mathrm{FBy}(z) \cdot e(z) + C_\mathrm{FBu}(z) \cdot u_\mathrm{lim}(z)
    .
    \label{eqn:ADRC_DT_DFTF_ulim}
\end{equation}

However, a remedy can be found by enhancing the feedback path through the observer represented by $ C_\mathrm{FBu}(z)$.
If (at time $k$) one does not only feed back $u_\mathrm{lim}(k)$, but additionally the part of the control signal that exceeded the limitation, i.e.\, $u(k) - u_\mathrm{lim}(k)$, this feedback loop will not ``forget'' the original value $u(k)$ computed by the control law:
\begin{equation}
    u(z) = C_\mathrm{FBy}(z) \cdot e(z) + C_\mathrm{FBu}(z)
    \cdot \Big[ u_\mathrm{lim}(z) +
    \underbrace{ \left( u(z) - u_\mathrm{lim}(z) \right) }_{ \text{additional feedback} }
    \Big]
    .
    \label{eqn:ADRC_DT_DFTF_ulim_mod1}
\end{equation}

While \refEq{eqn:ADRC_DT_DFTF_ulim_mod1} could be trivially reduced to the unlimited case \refEq{eqn:ADRC_DT_DFTF}, let us leave it in this (seemingly useless) form with the redundant terms $\pm u_\mathrm{lim}(z)$.
The reason for this is that the first $u_\mathrm{lim}(z)$ feedback can additionally be delayed by $D$ time steps to achieve delay compensation if necessary, as discussed before:
\begin{equation}
    u(z) = C_\mathrm{FBy}(z) \cdot e(z) + C_\mathrm{FBu}(z)
    \cdot \left[ z^{-D} \cdot u_\mathrm{lim}(z) + \left( u(z) - u_\mathrm{lim}(z) \right) \right]
    .
    \label{eqn:ADRC_DT_DFTF_ulim_mod2}
\end{equation}

        \subsubsection{Testing the Resulting Controller}
\begin{figure}
    \includegraphics[page=1]{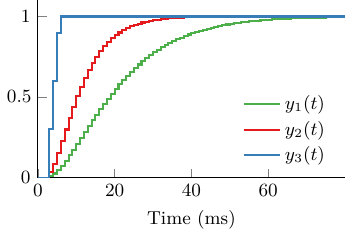}%
    \hfill%
    \includegraphics[page=2]{Demo_FilteredP}%
    \caption{%
        Simulation example for three different designs: controller output $u$ of the filtered P controller \refEq{eqn:ADRC_DT_DFTF_ulim_mod2} and controlled variable $y$ of an integrator plant \refEq{eqn:DT_IntegratorPlantDelay} with $b = b_0 = 10$ and $D = 2$ steps of time delay ($\Tsample = 1\,\mathrm{ms}$).
        Designed using \refEq{eqn:FilteredP_Design_Tsettle} for $T_\mathrm{settle} = 60\,\mathrm{ms}$ (\#1), $30\,\mathrm{ms}$ (\#2), and dead-beat behavior ($0\,\mathrm{ms}$, $\omega_\mathrm{ESO} \to \infty$, \#3) with an output limit $u_\mathrm{max} = 30$
    }
    \label{fig:Demo_FilteredP}
\end{figure}

%
%
reveals that we have already obtained a relatively powerful tool for its kind, although it still is a filtered instead of a pure P controller.
Especially when controlling an integrator plant, its discrete-time tuning equations not only allow to easily select the desired closed-loop dynamics, but also push the dynamics straight to the boundaries of dead-beat control---and in contrast to the latter, this controller is able to take time delays and output limitation into account ``out of the box'', as demonstrated in \refFig{fig:Demo_FilteredP}.

    \subsection{Removing the Filter}
    \label{sec:PureP}

On the way to parameterize discrete-time error-based ADRC to behave like a pure P controller, looking back at \refEq{eqn:FilteredP_CFB_Approach} suggests an easy solution: setting $\alpha_1^\mathrm{FB}$ in $C_\mathrm{FB}(z)$ to zero.
With bandwidth parameterization, \refEq{eqn:FilteredP_CFB_alpha1} specifies one has to set $z_\mathrm{ESO} = 0$ for that, which corresponds to an infinite observer bandwidth $\omega_\mathrm{ESO}$.
Summarizing the controller/observer gains and transfer function coefficients we obtain with the parameter choice $\omega_\mathrm{CL} = 0$ and $\omega_\mathrm{ESO} \to \infty$:
\begin{gather}
    k_1 = 0
    ,\quad
    l_1 = 1
    ,\quad
    l_2 = \frac{1}{\Tsample}
    ,\quad \text{as well as}
    \\
    \alpha_1^\mathrm{FB} = \alpha_1 = \alpha_2 = 0
    ,\quad
    \beta_0 = \frac{ 1 }{ b_0 \Tsample } = -\beta_1
    ,\quad
    \gamma_0 = 1
    ,\quad
    \gamma_1 = 0
    .
    \label{eqn:PureP_CFB_Coeffs}
\end{gather}

        \subsubsection{Open-Loop Properties}

in the linear case, where the controller output limitation is not active ($u_\mathrm{lim} = u$), now reveal a transfer function \refEq{eqn:PureP_TF} that gets simplified to a pure proportional controller.
To achieve a desired proportional controller gain $K = \frac{ 1 }{ b_0 \Tsample }$ one therefore has to select a parameter value $b_0 = \frac{ 1 }{ K \Tsample }$.

\begin{equation}
    C_\mathrm{FB}(z)
    = \frac{ 1 }{ b_0 \Tsample } \cdot \frac{ 1 - z^{-1} }{ 1 - z^{-1} }
    \stackrel{{\color{orange}u_\mathrm{lim} = u}}{=}
    \frac{ 1 }{ b_0 \Tsample }
    .
    \label{eqn:PureP_TF}
\end{equation}

        \subsubsection{Closed-Loop Dynamics with an Integrator Plant}
%
%
$P(z)$ from \refEq{eqn:DT_IntegratorPlant} will now be examined for the two cases without and with perfect knowledge $b_0 = b$:
\begin{equation}
    G_\mathrm{CL}(z)
    = \frac{ C_\mathrm{FB}(z) \cdot P(z) }{ 1 + C_\mathrm{FB}(z) \cdot P(z) }
    = \frac{ \frac{b}{b_0} \cdot z^{-1} }{ 1 + \left( \frac{b}{b_0} - 1 \right) \cdot z^{-1} }
    \stackrel{{\color{orange}b_0 = b}}{=}
    z^{-1}
    .
    \label{eqn:PureP_GCL_Integrator}
\end{equation}

Comparison with generic first-order low-pass behavior in \refEq{eqn:DT_LowPass} allows to derive a tuning equation to determine the single parameter $b_0$ which is now left depending on a desired bandwidth $\omega_\mathrm{LP}$ of the closed-loop low-pass dynamics:
\begin{equation}
    b_0 = \frac{ b }{ 1 - \mathrm{e}^{-\omega_\mathrm{LP} \Tsample } }
    .
\end{equation}

For $b_0 = b$ the closed-loop transfer function $G_\mathrm{CL}(z)$ is simplified and yields dead-beat behavior for the controlled integrator plant.
While we did not aim at this in the first place, it does not come as a surprise now, since the closed-loop transfer function \refEq{eqn:FilteredP_GCL_Integrator} with the filtered P variant from \refSec{sec:FilteredP} will converge to dead-beat dynamics for increasing bandwidth values $\omega_\mathrm{ESO} \to \infty$ (or $z_\mathrm{ESO} \to 0$).

        \subsubsection{Delay Compensation}
%
%
is now performed as in the filtered variant by delaying the feedback path to the observer (or rather, what is left of it).
Considering the coefficients in \refEq{eqn:PureP_CFB_Coeffs}, the ``dual feedback'' transfer functions are reduced to $C_{\mathrm{FBy}}(z) = \beta_0 \cdot \left( 1 - z^{-1} \right)$ and $C_{\mathrm{FBu}}(z) = \gamma_0 \cdot z^{-1}$, therefore we obtain the controller transfer function in the linear (unlimited) case as \refEq{eqn:PureP_CFB_TF_DelayComp}.
With the example of the integrator plant $P(z)$ \refEq{eqn:DT_IntegratorPlant}, the closed-loop transfer function becomes \refEq{eqn:PureP_GCL_IntegratorDelay}.
\begin{equation}
    C_\mathrm{FB}(z)
    = \frac{ C_\mathrm{FBy}(z) }{ 1 - C_\mathrm{FBu}(z) \cdot z^{-D} }
    = \frac{ 1 }{ b_0 \Tsample } \cdot \frac{ 1 - z^{-1} }{ 1 - z^{-1-D} }
    .
    \label{eqn:PureP_CFB_TF_DelayComp}
\end{equation}
\begin{equation}
    G_\mathrm{CL}(z)
    = \frac{ C_\mathrm{FB}(z) \cdot P(z) }{ 1 + C_\mathrm{FB}(z) \cdot P(z) }
    = \frac{ \frac{b}{b_0} \cdot z^{-1-D} }{ 1 + \left( \frac{b}{b_0} - 1 \right) \cdot z^{-1-D} }
    \stackrel{{\color{orange}b_0 = b}}{=}
    z^{-1-D}
    .
    \label{eqn:PureP_GCL_IntegratorDelay}
\end{equation}

The good news from \refEq{eqn:PureP_GCL_IntegratorDelay} is that for $b_0 = b$ dead-beat dynamics are obtained again, with the fastest possible response considering the additional loop delay $z^{-D}$.
The bad news concerns all other cases $b_0 \ne b$.
While $G_\mathrm{CL}(z)$ looks strikingly similar to the result without delay in \refEq{eqn:PureP_GCL_Integrator}, the response can be considered as sampled down by a factor of $(1+D)$.
This comes from the control signal, which will consist of a repeating sequence with one non-zero followed by $D$ zero-valued intervals, as one can verify from a step response of the reference-to-control-signal transfer function.
With an abbreviation $a = 1 - \frac{b}{b_0}$ the control signal is:

\begin{align}
    u_\mathrm{step}(z)
    &= \frac{ C_\mathrm{FB}(z) }{ 1 + C_\mathrm{FB}(z) \cdot P(z) } \cdot \frac{ 1 }{ 1 - z^{-1} }
    = \frac{1}{b_0 \Tsample} \cdot \frac{ 1 }{ 1 - a \cdot z^{-1-D} }
    \\
    &= \frac{1}{b_0 \Tsample} \cdot \left(
    a^0 + a^1 \cdot z^{-(1+D)} + a^2 \cdot z^{-2 \cdot (1+D)} + a^3 \cdot z^{-3 \cdot (1+D)} + \ldots
    \right)
    \notag
    .
\end{align}
As an example, for $D = 2$ the sequence of $u$ values is $\{a^0, 0, 0, a^1, 0, 0, a^2, 0, 0, \ldots\}$, clearly unacceptable behavior.
We therefore have to conclude that, unfortunately, the unfiltered P variant obtained by tuning ADRC with $\omega_\mathrm{CL} = 0$ and $\omega_\mathrm{ESO} \to \infty$ leaves us with a P controller that does not inherit the useful option of delay compensation from its filtered variant.
The remainder of this article is therefore devoted to learning from these observations and creating a P controller from a targeted reduction of discrete-time error-based ADRC to the essentials.

\section{Simplifying Error-Based ADRC to a P Controller}
\label{sec:FinalP}

    \subsection{Parameter Choice and Closed-Loop Dynamics}

Removing the filter by letting $\omega_\mathrm{ESO} \to \infty$ for $z_\mathrm{ESO} \to 0$ turned out to introduce new problems.
A better solution can be found by going only halfway, setting $k_1 = 0$ and $l_1 = 1$ as before in \refSec{sec:PureP}, but keeping $l_2$ variable.
The transfer function coefficients are then reduced to:
\begin{equation}
    \alpha_1^\mathrm{FB} = 0
    ,\ \
    \alpha_1 = l_2 \Tsample - 1
    ,\ \
    \alpha_2 = 0
    ,\ \
    \beta_0 = \frac{ l_2 }{ b_0 } = -\beta_1
    ,\ \
    \gamma_0 = l_2 \Tsample
    ,\ \
    \gamma_1 = 0
    ,
    \label{eqn:FinalP_CFB_Coeffs}
\end{equation}
resulting in the following feedback controller transfer function including delay compensation for $D$ steps:
\begin{equation}
    C_\mathrm{FB}(z)
    = \frac{ C_\mathrm{FBy}(z) }{ 1 - C_\mathrm{FBu}(z) \cdot z^{-D} }
    = \frac{ l_2 }{ b_0 } \cdot \frac{ 1 - z^{-1} }{ 1 + \left( l_2 \Tsample - 1 \right) \cdot z^{-1} - l_2 \Tsample \cdot z^{-1-D} }
    .
    \label{eqn:FinalP_CFB}
\end{equation}

Clearly $C_\mathrm{FB}(z)$ in \refEq{eqn:FinalP_CFB} is reduced to pure P controller behavior $C_\mathrm{FB}(z) = \frac{ l_2 }{ b_0 }$ when used without delay compensation ($D = 0$).
Closing the loop with the previously used delayed integrator plant $P(z)$ from \refEq{eqn:DT_IntegratorPlantDelay} yields:

\begin{equation}
    G_\mathrm{CL}(z)
    = \frac{ C_\mathrm{FB}(z) \cdot P(z) }{ 1 + C_\mathrm{FB}(z) \cdot P(z) }
    = \frac{ l_2 \Tsample \frac{b}{b_0} \cdot z^{-1-D} }{ 1 + \left( l_2 \Tsample - 1 \right) \cdot z^{-1} + l_2 \Tsample \cdot \left(\frac{b}{b_0} - 1 \right) \cdot z^{-1-D} }
    .
    \label{eqn:FinalP_GCL_IntegratorDelay}
\end{equation}

This result includes the previously achieved special cases of the closed-loop behavior for the options delay compensation, perfect knowledge $b_0 = b$, and dead-beat tuning for $l_2$:
\begin{equation}
    G_\mathrm{CL}(z)
    \stackrel{{\color{orange}b_0 = b}}{=}
    \frac{ l_2 \Tsample \cdot z^{-1-D} }{ 1 + \left( l_2 \Tsample - 1 \right) \cdot z^{-1} }
    \stackrel{{\color{orange}l_2 = \frac{1}{\Tsample}}}{=}
    z^{-1-D}
    .
\end{equation}
\begin{equation}
    G_\mathrm{CL}(z)
    \stackrel{{\color{orange}D = 0}}{=}
    \frac{ l_2 \Tsample \frac{b}{b_0} \cdot z^{-1} }{ 1 + \left( l_2 \Tsample \frac{b}{b_0} - 1 \right) \cdot z^{-1} }
    \stackrel{{\color{orange}b_0 = b}}{=}
    \frac{ l_2 \Tsample \cdot z^{-1} }{ 1 + \left( l_2 \Tsample - 1 \right) \cdot z^{-1} }
    \stackrel{{\color{orange}l_2 = \frac{1}{\Tsample}}}{=}
    z^{-1}
    .
\end{equation}

Apart from dead-beat tuning $l_2 = \frac{1}{\Tsample}$ the comparison with \refEq{eqn:DT_LowPass} results in a tuning equation if $b_0 = b$ is known (but be reminded that the settling time does not include the time delay):
\begin{equation}
    l_2 = \frac{  1 - \mathrm{e}^{-\omega_\mathrm{LP} \Tsample } }{ \Tsample }
    \quad \text{with} \quad
    \omega_\mathrm{LP} \approx \frac{ 4 }{ T_\mathrm{settle,98\%} }
    ,
    \label{eqn:FinalP_CFB_Design_Tsettle}
\end{equation}
which, not by accident, corresponds to the tuning equation \refEq{eqn:ADRC1_DT_KL_BWTuning} for $k_1$ of the outer loop of error-based ADRC, as $k_1$ is calculated for a unity-gain discrete-time integrator plant normally ensured by ADRC's disturbance rejection loop.

    \subsection{Simplified Implementation}

To keep the nice properties discussed in the filtered controller variant from \refSec{sec:FilteredP} (handling controller output limitation and delay compensation), the following modifications have to be applied to the $u$-feedback channel of the ``dual feedback'' transfer function structure:
\begin{itemize}
    \item
    apply a delay of $D$ steps to the value of $u$ fed back into the observer, and

    \item
    in case of an active controller output limitation (where $u_\mathrm{lim}$ is fed back instead of $u$), additionally feed back the non-realizable part $u - u_\mathrm{lim}$.
\end{itemize}
One then obtains from rearranging \refEq{eqn:ADRC_DT_DFTF_ulim_mod2}:
\begin{align}
    u(z) &=
    \frac{ C_\mathrm{FBy}(z) }{ 1 - C_\mathrm{FBu}(z) } \cdot e(z)
    + \frac{ C_\mathrm{FBu}(z) \cdot \left( -1 + z^{-D} \right) }{ 1 - C_\mathrm{FBu}(z) } \cdot u_\mathrm{lim}(z)
    \notag
    \\
    &=
    \frac{ l_2 }{ b_0 } \cdot e(z)
    + l_2 \Tsample \cdot \frac{ -z^{-1} + z^{-1-D} }{ 1 - z^{-1} } \cdot u_\mathrm{lim}(z)
    .
\end{align}

A very interesting simplification can be achieved by polynomial long division of the transfer function applied to the feedback path of the limited control output, or, alternatively, recognizing that this transfer function resembles that of a moving average filter of order $D$ multiplied with $-z^{-1}$:

\begin{equation}
    u(z)
    = \frac{ l_2 }{ b_0 } \cdot e(z)
    - l_2 \Tsample \cdot \left( \sum_{\nu = 1}^{D} z^{-\nu} \right) \cdot u_\mathrm{lim}(z)
    .
    \label{eqn:FinalP_CFB_Implementation}
\end{equation}

\begin{figure}[t]
    \includegraphics[page=1]{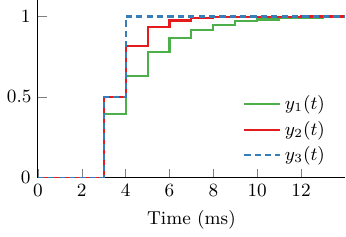}%
    \hfill%
    \includegraphics[page=2]{Demo_FinalP}%
    \caption{%
        Simulation example for three different designs: controller output $u$ of the final P controller \refEq{eqn:FinalP_CFB_Implementation} and controlled variable $y$ of an integrator plant \refEq{eqn:DT_IntegratorPlantDelay} with $b = b_0 = 10$ and $D = 2$ steps of time delay ($\Tsample = 1\,\mathrm{ms}$).
        Designed using \refEq{eqn:FinalP_CFB_Design_Tsettle} for $T_\mathrm{settle} = 8\,\mathrm{ms}$ (\#1), $4\,\mathrm{ms}$ (\#2), and dead-beat behavior ($l_2 = \frac{1}{\Tsample}$, \#3) with an output limit $u_\mathrm{max} = 50$
    }
    \label{fig:Demo_FinalP}
\end{figure}

\begin{figure}[t]
    \includegraphics{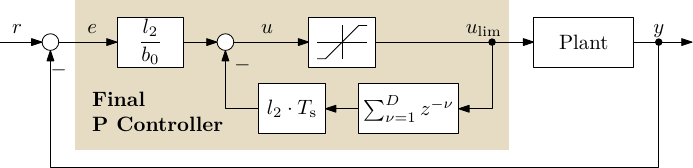}%
    \caption{%
        Control loop with the final, simplified implementation, with out-of-the-box capabilities of handling output limits and delay compensation.
        Dynamics up to dead-beat behavior when applied to an integrator plant $b/s$ can easily be achieved with the tuning equation \refEq{eqn:FinalP_CFB_Design_Tsettle} when its gain factor is known ($b_0 = b$)
    }
    \label{fig:FinalP_Implementation}
\end{figure}

The final result \refEq{eqn:FinalP_CFB_Implementation} is all that is left from simplifying error-based ADRC tuned for P controller behavior.
The successful integration of the practical features delay compensation and controller saturation are demonstrated in \refFig{fig:Demo_FinalP}.
As visible in \refFig{fig:FinalP_Implementation}, this controller is very trivial to implement.
For $D = 1$, for example, the implementation only needs two multiplications and one unit delay.

    \subsection{Equivalence to a Smith Predictor}

When applying the controller to the delayed integrator plant \refEq{eqn:DT_IntegratorPlantDelay}, a (maybe unexpected) equivalence to a Smith predictor can be proven.
A Smith predictor structure would, in the present case, replace the controller by a feedback loop $C_\mathrm{SP}(z)$ consisting of the P controller $C_\mathrm{FB}(z) = \frac{l_2}{b_0}$, the non-delayed plant model $P(z)$ from \refEq{eqn:DT_IntegratorPlant}, and a comb filter $1 - z^{-D}$:
\begin{align}
    C_\mathrm{SP}(z)
    &= \frac{ u(z) }{ e(z) }
    = \frac{ C_\mathrm{FB}(z) }{ 1 + C_\mathrm{FB}(z) \cdot P(z) \cdot \left( 1 - z^{-D} \right) }
    = \frac{
        \frac{l_2}{b_0}
    }{
        1 + \frac{l_2}{b_0} \cdot \frac{ b_0 \cdot \Tsample \cdot z^{-1} }{ 1 - z^{-1} } \cdot \left( 1 - z^{-D} \right)
    }
    \notag
    \\
    &= \frac{
        \frac{l_2}{b_0}
    }{
        1 + l_2 \Tsample \cdot \frac{ z^{-1} - z^{-1-D} }{ 1 - z^{-1} }
    }
    = \frac{
        \frac{l_2}{b_0}
    }{
        1 + l_2 \Tsample \cdot \sum_{\nu = 1}^{D} z^{-\nu}
    }
    ,
\end{align}
which corresponds to \refEq{eqn:FinalP_CFB_Implementation} for the linear (unlimited) case.
A clear advantage of \refEq{eqn:FinalP_CFB_Implementation} over the Smith predictor structure is its simplified implementation requiring only $D$ unit delays, as opposed to the plant model plus comb filter.

\section{Conclusion}

An existing error-based ADRC implementation, parameterized for $\omega_\mathrm{CL} = 0$, can be used to mimic the behavior of a P controller---when $\omega_\mathrm{ESO} < \infty$, with an additional low-pass filter.
Applied to an integrator plant, the desired observer dynamics will determine the closed-loop behavior, which makes tuning especially trivial, as the observer tuning equations from ADRC can be reused.
While delay compensation is easily possible in that case, graceful handling of controller output limitations requires an extension to the observer feedback path.
With $\omega_\mathrm{ESO} \to \infty$ the controller exhibits pure (unfiltered) P behavior, but does not get along well with ADRC's delay compensation approach.

Overcoming these limitation, as the final result of this article, a discrete-time proportional controller \refEq{eqn:FinalP_CFB_Implementation} with adjustable proportional gain was derived from parameterizing and then simplifying error-based ADRC of order one.
In contrast to a simple P controller, it can be tuned for dead-beat behavior in the closed loop with an integrator plant even when (known) delays are present in the loop.
With delay compensation for $D$ steps, this controller is equivalent to a P controller with Smith predictor designed for a discrete-time integrator plant, but with a simpler implementation.
When limiting the controller output, the controller behaves well and achieves the closest possible to the desired dynamics, even when aiming for a dead-beat design.


\end{document}